# A Study on Performance Analysis Tools for Applications Running on Large Distributed Systems

Ajanta De Sarkar [1] and Nandini Mukherjee [2]

[1] Department of Computer Application,
Heritage Institute of Technology,
West Bengal University of Technology, Kolkata 700 107, India
adesarkar@yahoo.com

[2] Department of Computer Science and Engineering,
Jadavpur University, Kolkata 700 032, India
nmukherjee@cse.jdvu.ac.in

**Abstract.** The evolution of distributed architectures and programming paradigms for performance-oriented program development, challenge the state-of-the-art technology for performance tools. The area of high performance computing is rapidly expanding from single parallel systems to clusters and grids of heterogeneous sequential and parallel systems. Performance analysis and tuning applications is becoming crucial because it is hardly possible to otherwise achieve the optimum performance of any application. The objective of this paper is to study the state-of-the-art technology of the existing performance tools for distributed systems. The paper surveys some representative tools from different aspects in order to highlight the approaches and technologies used by them.

## 1. Introduction

During the last decade, automatic or semi-automatic performance tuning of applications running on parallel systems has become one of the highly focused research areas. Performance tuning means to speed up the execution performance of an application. Performance analysis plays an important role in this context, because only from performance analysis data the performance properties of the application can be observed. A cyclic and feedback guided methodology is often adopted for optimizing the execution performance of an application – every time the execution performance of the application is analyzed and tuned and the application is then sent back for execution again in order to detect the existence of further performance problems. Thus, appropriate and detailed performance data captured during the execution of an application may effectively be used to tune the performance of an application.

Performance analysis tools are based on an event model of the execution. The events can be low-level events, such as a cache miss or the graduation of a floating-point instruction, as well as high-level events, such as start and end of a user function. Performance analysis tools depend on the information gathered for events during the execution. The information can be merely the existence of an event. More detailed information can also be gathered for more sophisticated analysis of the performance. The information can either be gathered in the form of summaries or individually for each event. Performance analysis in this context means getting relevant performance information from trace data. Thus the scope of performance monitoring and analysis comprises instrumentation (to decide which data is relevant in a particular context), measurement (or monitoring these data), data reduction and correlation, analysis and presentation (to identify and locate the performance problems) and finally, optimization.

In distributed environments like Grid, heterogeneous resources are connected through heterogeneous types of networks. Monitoring such an environment for executing compute-intensive applications and tuning the performance of these applications are major challenges faced by today's scientific community. The heterogeneous features of the computational resources, potentially unreliable networks connecting these resources and different administrative management domains complicate the issues related to improving performance of these environments and the performance of applications running in these environments. Often the post-mortem analysis and performance tuning on an incremental basis is unsuitable in such environments. For successful performance tuning, real time performance data needs to be captured. The requirements for successful performance tuning in a heterogeneous distributed environment are:
- to capture top to bottom and end to end real-time performance data;
- to apply state-of-the art technologies of the performance tools to correlate and interpret the captured data;
- dynamic policy selection in response to current resource availability and application demands.

The objective of this paper is to study the state-of-the-art techniques used by the existing performance analysis tools and to focus on the appropriate techniques for performance analysis of heterogeneous distributed environments that can effectively be used for application performance improvement in real-time. This paper focuses on some representative performance analysis tools and highlights the technologies used by them. The rest of this paper is organized as follows: Section 2 presents overview of the performance tools. As case studies, we discuss only five (*SCALEA*, *SCALEA-G*, *AKSUM*, *Pablo*, *EXPERT*) of the existing tools. Sections 3, 4, 5 and 6 focus on different techniques used by these tools for instrumenting the applications, measurement and data collection, performance analysis and analysis data presentation. A comparative study of all these tools is presented in Section 7. Future work is outlined in Section 8 followed by conclusion.

## 2. Existing Performance Analysis Tools

Several tools for measuring or analyzing performance of serial / parallel programs have been developed so far. These tools were designed with diverse objectives, targeted different parallel architectures and adopted various techniques for collection and analysis of performance data. The scope of these tools comprises instrumentation, measurement (monitoring), data reduction and correlation, analysis and presentation and finally, in some cases, optimization.

This paper focuses on the state of the art technologies used in the existing performance analysis tools. We studied some of the existing tools, namely *SCALEA*, *SCALEA-G*, *AKSUM*, *Pablo*, and *EXPERT*. In this section an overview of each of these performance analysis tools is presented. Following sections discuss the technologies used by these tools in more detail.

## *SCALEA*

*SCALEA* [1] is a performance instrumentation, measurement, and analysis system for distributed and parallel architectures that currently focuses on OpenMP, MPI, HPF programs, and mixed programming paradigms such as OpenMP / MPI. *SCALEA* compares and evaluates the outcome of several performance measurement experiments, which are carried out in order to analyze the performance of an application. The architecture of *SCALEA* consists of the following components
- *SCALEA* Instrumentation System (SIS)
- *SCALEA* Runtime System
- *SCALEA* Performance Data Repository
- *SCALEA* Performance Analysis & Visualization System

The function of each component is discussed in the following sections. Each component can be used by external tools as they provide open interfaces.

In *SCALEA,* the data repository is in PostgreSQL. *SCALEA* also supports multiple experiment performance analysis and the user can select several experiments and performance metrics of interest whose associated data are stored in the data repository.

## *SCALEA-G*

*SCALEA-G* [2] is a unified system for monitoring and analyzing performance in Grid environment. *SCALEA-G* is based on the concept of Grid Monitoring Architecture (GMA) [6] as proposed by Global Grid Forum and is implemented as part of Open Grid Services Architecture (OGSA) [7] based services. It provides an infrastructure of OGSA-compliant grid services for online monitoring and performance analysis of a variety of Grid services including computational resources, networks, and applications. It offers services like Directory Service, Archival Service, Sensor Manager Service, Source Code Instrumentation Service, Mutator Service, Client Service and Register Service. Besides these services, the other components of *SCALEA-G* are Graphical User Interface, Performance Analyzer, Instrumentation Mediator, Grid Experiment Data Repository, Sensor Repository, System Sensors and Application Sensors. Both push and pull models are supported for gathering monitoring data, thereby providing flexible and scalable monitoring and performance analysis.

## *AKSUM*

*Aksum* [3] is a multi-experiment performance analysis tool. This is a part of Askalon project. *Aksum* supports message passing, shared memory and mixed parallelism programs written in Fortran. It also analyzes Java codes.

*Aksum*'s architecture [11] includes the following components:
- User Portal
- Search Engine
- Instrumentation Engine
- Experiment Generator

- Experiment Launcher
- Experiment Data Repository

The performance metrics are defined in terms Performance properties [9] and are defined in JavaPSL. These may be freely edited, removed from or added to *Aksum* in order to customize and speedup the search process. *Aksum* can automatically:
- instrument the user's application;
- generate versions of the application using a set of user-supplied input parameters;
- collect the data generated by the instrumentation and analyze it; and
- relate the performance problems back to the source code and compare the performance behavior across multiple experiments.

*Pablo toolkit*

The goal of Pablo project [4] has been the development of a portable performance data analysis environment that can be used with a variety of massively parallel systems. The solution to Performance Optimization problem as pointed out by the Pablo group [14] is "integration of dynamic performance instrumentation and on-the-fly performance data reduction with configurable, malleable resource management algorithms and a real-time adaptive control mechanism". Pablo provides a performance analysis environment designed for performance data capture, analysis and presentation across a wide variety of scalable parallel systems.

A set of tools and libraries are provided by Pablo Group for performance data collection, analysis and presentation. The Pablo Group has also investigated techniques, which are needed to integrate dynamic performance data with information on program transformations.

*EXPERT*

The *EXPERT* [5] tool environment provides a complete tracing-based solution for automatic performance analysis of MPI, OpenMP, or hybrid applications running on SMP cluster machines. The *EXPERT* architecture consists of the following three components:
- The *EXPERT* performance tool (Extensible PERformance Tool)
- The *EARL* trace analysis language (Event Analysis and Recognition Language)
- The *EPILOG* tracing library (Event Processing, Investigating and LOGging)

*EXPERT* describes performance problems using a high level of abstraction in terms of common situations that result from an inefficient use of the underlying programming models. The analysis is carried out along three interconnected dimensions: class of performance behavior, call tree position, and thread of execution.

With the above introduction of the five representative tools, we now concentrate on the detailed study of each of these tools. The following four sections review the different techniques used by these performance analysis tools for traditional parallel and distributed architectures. The discussion mainly focuses on four different aspects of performance analysis tools, namely *instrumentation, measurement and data collection, performance analysis*, and *analysis data presentation*.

### 3. Instrumentation of Applications

This section discusses the instrumentation techniques used by the five existing performance analysis tools studied in this paper.

*SCALEA* uses the instrumentation system SIS (SCALEA Instrumentation System) to allow users to instrument their program. The system supports instrumentation of OpenMP, MPI, HPF and hybrid parallel programs (e.g. OpenMP / MPI). Programs can be instrumented in two modes: either using command line options or using directives. Invoking SIS program with command line options let the users to specify predefined code regions and performance metrics and analyze the entire program. The directives also allow the users to specify code regions and performance metrics. SIS will then instrument and analyze the program based on the inserted directives. Using SIS directives the programmer can also enable / disable instrumentation process.

SIS generates an instrumentation description file and stores that in a repository to relate all gathered performance data back to the input program in a later time. SIS is based on the instrumentation libraries SISPROFILING and PAPI. SISPROFILING supports profiling with times and the interface with PAPI library determines the hardware parameters.

SIS offers an interface for other tools to traverse the Abstract Syntax Tree (AST) representation of the code in order to specify code regions for which performance metrics should be obtained.

In *SCALEA-G,* the instrumentation of application can be done at source code level by using Source Code Instrumentation Service or dynamically at runtime through Mutator Service.

In the first approach, a Source Code Instrumentation Service (SCIS) is implemented, which is based on SCALEA Instrumentation System [1]. The SCIS only instruments input source files in Fortran, and the client has to compile and link the instrumented files with the measurement library containing application sensors.

In the second approach, the dynamic instrumentation mechanism is based on Dyninst. A mutator service is implemented as a GSI-based SOAP C++ Web service. An XML-based instrumentation request language (IRL) is developed to allow the client to specify code regions of which performance metrics should be determined. Hence the user can control the instrumentation process.

*Aksum* provides a user portal to enable the user to input information about the application, the target machine and the performance properties which are of interest. A search engine, after obtaining the user input, automatically selects and instruments code regions for collecting raw performance data based on which performance properties are computed. Instrumentation of the application code is done through Instrumentation Engine. The Instrumentation Engine of *Aksum* provides an interface to SCALEA and is used for instrumenting Fortran applications. On the other hand, Twilight is used for instrumenting Java programs. Twilight can statically instrument Java source codes and dynamically instrument Java class files.

The *Pablo* software instrumentation system has two subcomponents, an incrementing parser and associated graphical user interface and a performance data capture library. These components interchange their information through SDDF (Self-Defining Data Format) data format. The SvPablo browser of *Pablo toolkit* provides a graphical user interface for instrumenting source code. Applications can be instrumented either interactively or automatically.

For explicit message passing codes, the graphical instrumentation interface allows users to interactively specify either the local or global instrumentation. SvPablo parses each source file and flags constructs (outer loops and function or subroutine calls) that can be instrumented. The user then selects the events to be instrumented and SvPablo generates a new version of the source code containing instrumentations calls. Alternatively, the SvPablo stand-alone parser can be used to instrument all subroutines and outer loops, including MPI calls. Then the instrumented source code can be compiled and linked with the data capture library.

The instrumentation software supports tracing, interval timing, and counting. In all cases, the instrumentation library monitors the instrumentation overhead and generated data volume. If the overhead volume exceeds the high water mark, then statistical clustering can further reduce the data volume.

In *EXPERT* the user can generate event traces for C, C++, and Fortran applications just by linking to the EPILOG (Event Processing Investigating and LOGging). The EPILOG library is used to generate event traces from parallel applications including MPI, OpenMP, or hybrid applications distributed across one or more clusters of SMP nodes.

### 4. Measurement and Data Collection

This section discusses the measurement and data collection mechanisms in each of the five tools.

In *SCALEA*, the raw performance data are collected for post-mortem or online analysis and stored in the performance data repository. The *SCALEA* performance data repository holds relevant information about the experiments conducted. Each experiment is described by experiment-related data, which includes information about the application code, the part of a machine on which the code has been executed, and performance information. An application (program) may have a number of implementations (code versions), each of which consists of a set of source files and is associated with one or several experiments.

*SCALEA* provides a performance database that is used to store all performance data, overheads, source codes, as well as relevant data of users' programs. Currently performance database is built on PostgreSQL.

*SCALEA-G*, which is based on the Grid Monitoring Architecture, collects performance data using different kinds of sensors. In *SCALEA-G*, two types of sensors are used – system sensors and application sensors. These sensors generate time-stamped performance monitoring events for hosts, network processes and applications. A Sensor Manager is responsible for starting and stopping the sensors. A directory service is maintained for publication of the locations of all sensors and other event suppliers. Producers keep the sensor directory up-to-date and listen to data requests from event consumers.

A sensor repository is used to hold the information about available sensors. XML schema is used to express sensors in the sensor repository. The XML schema allows specifying sensor-related information such as name (a unique name of the sensor), measureclass (implementation class), schema-file (XML schema of data produced by the sensor), parameters (which are required for invoking the sensor) etc. The monitored data is propagated via a data channel. An XML schema describes each kind of monitored data. So any client can easily access the data via XPath/XQuery.

As proposed in GMA, *SCALEA-G* supports both push and pull models. During the flexible and scalable performance monitoring, data subscription is done through push model and data query is performed through query model. In push model, any new data entry will be sent to the subscribed consumers after checking the subscribed condition. In pull model, Sensor Manager Service only searches current available entries in the data buffer and entries met conditions of consumer query will be returned to the requested consumers. Consumer builds pull/push request in XML based XPath/XQuery.

As in *SCALEA*, *Aksum* collects the performance data by generating and executing some experiments. For each instrumented application an experiment generator decides the experiments to be carried out by setting the input files, the number of threads or processes to be used for the application execution and so on. An experiment launcher compiles and executes the application instance on the target machine.

*Aksum* assumes that the experiment terminated normally if performance data has been stored in the experiment data repository. *Aksum* enables the user to select two additional options for terminating experiments:
- the execution time of an application instance exceeds a user-defined time value.
- After polling the CPU n times every k seconds, *Aksum* detects that the load of every CPU used by a given experiment is always below a certain value f (where n, k and f are user-defined values).

In **Pablo toolkit**, the Pablo Performance Capture Facility (PCF) supports MPI and provides several options for gathering time stamped event records and extensions for recording performance information for MPI calls, Unix I/O, and MPI I/O operations. The library silently records the occurrence of each event to trace files in the SDDF format. The Pablo data capture library also monitors the aggregate event rate. This library has been designed to reduce the likelihood of malignant perturbations by monitoring and dynamically altering the volume, frequency, and types of event data recorded.

In *Pablo toolkit*, for both message passing and data parallel codes, the data capture software can compute summaries of the performance data, or the data can be recorded for subsequent, off-line analysis.

Performance data is gathered at runtime and stored in a performance file. This file is stored in the Pablo Self-Defining Data Format (SDDF) and this is input to SvPablo. The SDDF metaformat is useful to represent performance data flexibly. SDDF is a data description language that specifies both data record structures and data record instances. SDDF has both a compact library version and a human-readable ASCII version.

In **EXPERT**, EPILOG trace library is used for data collection. The event traces generated by executing parallel applications are stored in trace files in EPILOG binary trace data format. The trace files are used as input to the *EXPERT* performance tool. The library maps locations of the events within the hardware architecture, their processes, and threads of execution. The performance data includes all necessary source code and call site information, hardware performance counter values, and marking of collectively executed operations for both MPI and OpenMP.

The EARL trace analysis language (Event Analysis and Recognition Language) maps an EPILOG trace file to the EARL event trace model, which provides abstractions in order to describe the compound events representing inefficient behavior. Depending on the event type, each event is characterized by a set of attributes. The event types are organized in a hierarchy. The event type itself can also be accessed as value of an attribute.

EARL provides two types of abstractions- System states and Pointer attributes. System states map individual events onto a set of events that represent one aspect of the parallel system's execution state at the moment when the event happens. Pointer attributes connect related events, so that compound events can be defined along a path of related events.

### 5. Performance Analysis

Performance Analyzer is the most important component of any performance tool, because it enables the user to interpret the huge volume of monitoring data in terms of performance problems in specific code regions.

The *SCALEA* performance analysis module analyzes the raw performance data and computes all user-requested performance metrics for visualizing them together with the input program. The performance data are analyzed or extracted to XML form by *SCALEA* utilities. Performance Analysis reflects the dynamic relationship between dynamic code regions and their sub regions. It also provides a detailed overhead analysis for every code region.

*SCALEA* focuses on five major categories of temporal overheads and further classifies them in a hierarchical structure. The overhead categories are listed below:
- *Data movement* overhead that corresponds to local and remote accesses;
- *Synchronization* overhead due to barriers and locks;
- *Control of parallelism* overhead (time required for fork and join or loop scheduling)
- *Additional computation* overhead caused by algorithmic or compiler changes to increase parallelism or data locality.
- *Loss of parallelism* (due to imperfect parallelisation)

*SCALEA* also supports multi experiment performance analysis. Performance metrics are computed by analyzing data obtained from several experiments.

*SCALEA-G* [13] supports performance monitoring and analysis for Grid by providing instrumentation and sensor-based monitoring middleware. Analysis Control decides which activities should be instrumented, monitored and analyzed. Based on information about the selected activity instance and its consumed resources, performance analyzer analyzes collected monitoring data.

Analysis of any application is processed for two different models: fork-join model and multi-workflow model. The metrics like load imbalance, slowdown factor and synchronization delay can be measured using the fork-join model of work-flow activities and the impact of the slower activities on the overall performance of the whole structure can be analyzed.

Performance data are collected and analyzed at two levels: activity and workflow level. At activity level, performance metrics such as execution status, wall-clock time and hardware metrics of instrumented code regions are captured. Activity overhead analysis for overheads like communication and synchronization is done for each activity. At workflow level, performance metrics that characterize the interaction and performance impact among activities are monitored and analyzed. In the analysis phase, load imbalance, computation to communication ratio, activity usage, and success rate of activity invocation, average response time, waiting time, synchronization delay, etc. are computed.

*Aksum*, automatically selects and instruments code regions for collecting raw performance data based on which performance properties are computed. Performance properties (e.g. load imbalance, synchronization overhead) are hierarchically organized into tree structures called property hierarchies. Each node in the property hierarchy represents a performance property and is described by three elements: *Performance property name*, *Threshold* and *Reference code region*.

*Aksum* provides three standard property hierarchies involving message passing, shared memory and mixed parallel programs. The user can define and store new property hierarchies from scratch or customize predefined hierarchies. The search engine evaluates the performance data and tries to determine all critical performance properties. A cycle consisting of consecutive phases of application execution and property evaluation is continued until all experiments are done. *Aksum* assumes that the experiment terminated normally if performance data has been stored in the experiment data repository.

*Pablo* performance analysis environment [8] provides two different types of analysis:
- statistical analysis
- input/output analysis

In statistical analysis, histograms of data values are computed and displayed from any SDDF file. The generality of the SDDF infrastructure makes it possible to reuse the same analysis software as new performance data records are defined and added to the instrumentation suite. The statistical analysis program is distributed as part of the SDDF library component.

For input/output analysis, text-based reporting programs are used to study the effects of input/output statements on performance. The input/output performance of multiple versions of the same program or different programs can be compared with the help of this tool.

*EXPERT* uses EARL to map the event trace onto a higher level of abstraction to make the analysis process simple and easy to extend. Actually, the EARL trace analysis language (Event Analysis and Recognition Language) is one of the components of *EXPERT* tool. The analysis process is carried out by the *EXPERT* component.

At the time of analysis, performance properties of a parallel application are classified according to their influence on the performance of the application. A performance property [9] characterizes a class of performance behavior. A severity measure indicates the influence of a performance property for one run of an application. The performance properties are specified in form of patterns and can detect compound events indicating inefficient behavior.

*EXPERT* organizes the performance properties in a hierarchy. The upper levels of the hierarchy correspond to more general behavioral aspects such as time spent in MPI functions. The deeper levels correspond to more specific situations such as time lost due to blocking communication.

The representation of Performance behavior is done using a three-dimensional matrix; where each cell contains the severity for a specific performance property, call tree node, and location. The first dimension describes the kind of inefficient behavior. The second dimension describes both its source code location and the execution phase during which it occurs. Finally the third dimension gives information on the distribution of performance losses across different processes or threads.

## 6. User Interface and Presentation of Analysis Data

After collection and analysis of performance data, it must be related to the code sections and displayed in order to enable the user and/or tuning tools to take appropriate tuning actions.

In *SCALEA,* analysis data is presented through the visualization module. In this performance tool, performance analysis module stores the performance data in a performance repository. A graphical user interface is provided to retrieve data from the repository and view a large variety of performance metrics at the level of arbitrary code regions, threads, process, and computational nodes for single-and multi-experiments.

The module computes all user-requested performance metrics, and visualizes them together with the input program. Two modes of analysis in single experiment (Region-to-overhead, Overhead-to-Region) are visualized through the GUI, which allows the programmer to select any code region instance in the dynamic code region call graph. In case of multiple experiment analysis, the outcome of every selected metric is analyzed and visualized for all experiments.

In *SCALEA-G*, the user GUI uses Askalon Visualization, JFreeGraph for presentation of performance and monitoring data. The GUI is also used for administration and dynamic instrumentation. In the administration GUI, user can select a Sensor Manager Service, a list of available sensors and a list of sensor instances managed by that Sensor Manager Service. Also an existing sensor instance can be deactivated by selecting deactivate button. Detailed information of the sensor can also be displayed by choosing a sensor. In dynamic instrumentation GUI, the user can choose a directory service and retrieve a list of instances of Mutator Service registered to that directory service. The user can monitor processes running on computer nodes.

*Aksum* presents the analysis data through a graphical interface. A User Portal is used through which user can also input data. The User Portal displays only the property names for those instances whose severity is above the user-defined threshold. Expanding each propertyname can show the property instances. *Aksum* allows the user to modify and add new property and property hierarchy. This can be displayed in property visualization [12]. *Aksum* examines the performance for multi-experiments based on an arbitrary user-defined set of machine and problem size. A graph can be plotted showing the severity of the performance property for several experiments versus machine sizes.

In *Pablo toolkit*, output data is presented through graphical data flow metaphor. The data presentation techniques [11] used in Pablo are Dynamic Graphics, Head-Mounted Displays and sonification. In Dynamic graphics display, a variety of X windows performance data displays are developed including dials, bar charts, X-Y plots etc. along with a general-purpose graph display. Each display supports standard interface conventions, which promotes ease of use and rapid construction of new displays. Head-mounted displays provide a non-intrusive alternative to visual monitoring on a workstation. In sonification display, sonic widgets support the mapping of a sequence of scalar data values to musical scales with different tunings, volumes, timbers and arbitrary sound samples. *SvPablo* and *Virtue* are generally used as data presentation tools. *SvPablo*, a performance browser provides a hierarchy of color-coded performance displays, including a high-level routine profile and source code scroll boxes. *Virtue* provides a high modality performance display and analysis system. Through this display, users can interactively obtain high-level geographic views of network traffic, as well as can view low-level tasks.

*EXPERT* provides a unique display technique, weighted tree which is specially developed for output data presentation. A weighted tree is a tree browser that labels each node with a weight. EXPERT uses a percentage of the application's total CPU allocation time as weight (e.g., the percentage of time spent in a sub tree of the call tree). The weight is displayed simultaneously using both a numerical value as well as a colored icon. The weighted trees of the different analysis dimensions are interconnected, so that the user can display the call tree with respect to a particular performance property, and the distribution across the locations with respect to a particular node in call tree.

*EXPERT* is tested for two types of applications, pure MPI application and hybrid MPI / OpenMP applications on SMP cluster testbed. In both types of applications, visualization is based on performance properties, dynamic call tree and code locations. The performance properties help to identify actual problem, whereas dynamic call trees or region trees display where in the source code the problem occurs and code location displays how the problem is distributed across the machine. The values and colors for performance property represent percentages of the total CPU allocation time. The percentages in the dynamic call tree refer only to the selection of specific performance property and the percentages for locations refer only the specific process (thread) of execution.

## 7. A Comparative Analysis of the Tools

In this section we present a table (Table 1) showing comparisons among the tools described in the previous sections. We mainly concentrate on certain features like *instrumentation, data exchange format, types of analysis data* etc. The table demonstrates that although the tools offer similar kinds of services for performance analysis of application programs, the techniques used by them are diverse and apart from a few exceptional cases, no standard has yet been accepted. For distributed environments like Grid, a grid monitoring architecture is proposed. However, there is ample scope of further improvements and clarifications at different levels.

**Table 1 Comparative study of the five performance analysis tools**

| Features | SCALEA | SCALEA-G | AKSUM | Pablo | EXPERT |
|---|---|---|---|---|---|
| **Programming languages** | HPF, OpenMP, MPI and hybrid | Fortran | Fortran 90, MPI, OpenMP, hybrid, JAVA | C++, F77, F90 with MPI calls | MPI, OpenMP, hybnrid application running on SMP clusters |
| **Instrumentation** | With command-line options or directive-based using SISPROFILING and PAPI library | Directive-based source code instrumentation service and dynamic instrumentation using GSI-based SOAP C++ Web service | Automatic using SCALEA and Twilight. User has control through user portal | Automatic or interactive (using a GUI) source code instrumentation using SvPablo library | Using EPILOG trace library |
| **Data Capture** | Based on event tracing in (Single or Multiple) executions of the application | Producer-consumer model-based (consumer either subscribes (push) or queries (pull) for data) | Based on event tracing in (Single or Multiple) executions of the application | Based on event trace; volume, frequency and type of data can be dynamically altered | Based on event tracing |
| **Data storage** | PostgreSQL performance database | XML-based | Data repository is used | Data file | Data file |
| **Data Exchange Format** | XML | XML | JavaPSL | Self-Defining Data Format (SDDF) | EPILOG binary trace data format |
| **Performance Analysis** | Post-mortem, and online analysis | Online via data channel using tunnel protocol | Post-mortem, cyclic process of application execution and property evaluation | Post-mortem analysis using SvPablo [1] | Post-mortem using event driven approach |
| **Type of Analysis Data** | Hierarchical classification of overhead-based | Hierarchical classification of overhead-based | Performance Properties and their severity measures | Statistical analysis and input / output analysis | Performance Properties and their severity measures |
| **Analysis data presentation** | Graphical User Interface (GUI) | GUI using Askalon Visualization, JfreeGraph | User Portal (GUI) | Graphical data flow metaphor, Dynamic Graphics, Head-Mounted Displays, and Sonification | Weighted Tree |

---

[1] Autopilot supports real-time performance analysis and visualization, but this tool has not been considered in this paper.

## 8. Conclusion and Future Work

The purpose of this paper is to make a comparative study of some existing performance analysis tools and focus on the techniques used by them. Some of the points that have been observed throughout the above discussion are summarized below:

- The important components of the performance tools are instrumentation, measurement, analysis and presentation or visualization.
- Although the tools often make static instrumentation controlled by the users, dynamic instrumentation is more attractive and is a requirement for today's distributed computing environments like Grid.
- Many tools perform postmortem analysis based on data collected at runtime. However, due to the heterogeneous and dynamic nature of the today's distributed environment postmortem analysis is no longer possible.
- The tools use different data storage and exchange formats, which need to be standardized. In this regard, XML is a good option.
- Analysis data are differently presented by different tools. It is also required to standardize the types of analysis data and their presentation format. The APART working group [9] is working for some time to define the performance properties for describing various performance problems in applications.

Our future work focuses on developing performance analysis and tuning services for distributed environments. The essential ingredients of such real-time analysis and tuning services are

- *dynamic instrumentation and data collection mechanism* (as performance problems must be identified during run-time),
- *data reduction* (as the amount of monitoring data is large and movement of this amount of data through the network is undesirable),
- *low-cost data capturing mechanism* (as overhead due to instrumentation and profiling may contribute to the application performance), and
- *adaptability to heterogeneous environment* (for the heterogeneous nature of Grid, the components along with the communicating messages and data must have minimum dependence on the hardware platform, operating system and programming languages and paradigms).